\documentstyle[stwol,amsmath,cite,graphicx]{article}

\newcommand{\Eq}[1]{eq.~\ref{#1}}
\newcommand{\Ref}[1]{Ref.~\cite{#1}}
\newcommand{\Fig}[1]{Fig.~\ref{#1}} 

\begin{document}

\renewcommand{\thefootnote}{\fnsymbol{footnote}}

\title{HYPERON POLARIZATION AND VECTOR MESON PRODUCTION AT 2.85 GeV \\
            (Preliminary Results)}

\author{A. MAGGIORA} 

\address{Istituto Nazionale Fisica Nucleare, Via P. Giuria 1,\\
         10125 - Torino, Italy}

\twocolumn[\maketitle\abstracts{
           \begin{center} \mbox{\small On Behalf of the DISTO Collaboration                                   }~\footnotemark \end{center}
\vspace{4pt}
Surprisingly large polarization in hyperon production by unpolarized
proton beam has been known since long time. 
Huge inclusive hyperon polarization data are available in literature,
few data are disponible on spin observables and none on exclusive hyperon 
measurements. Evidence of relevant violation of the OZI rule in several
reactions are also shown in recent measurements.
These two items, connected together by the hypothesis of
existence of a sizeable $q \bar{q}$ sea in the nucleon, can be studied using
the DISTO is a spectrometer installed in the polarized proton beam
of the Saturne accelerator in Saclay.
The compact experimental set-up is designed to detect
four or more charged particles signaling $\Lambda$, $\Sigma^{0}$, $Y^{*}$
or $\phi$ production in $\vec{p} p$ interaction. 
An outline of the physical motivations and a brief description of the 
experimental apparatus is given as well as some preliminary results from 
the first production running.}]

\footnotetext{
     F. Balestra$^{d}$, Y. Bedfer$^{c}$, R. Bertini$^{c,d}$, L.C. Bland$^{b}$,
     A. Brenschede$^{h}$, F. Brochard$^{c}$, M.P. Bussa$^{d}$, S. Choi$^{b}$,
     M. Debowski$^{e}$, M. Dzemidzic$^{b}$, I.V. Falomkin$^{a}$,
     J.Cl. Faivre$^{c}$, L. Fava$^{d}$, L. Ferrero$^{d}$, J. Foryciarz$^{g}$, 
     V. Frolov$^{a}$, R. Garfagnini$^{d}$, D. Gill$^{l}$, A. Grasso$^{d}$, 
     E. Grosse$^{e}$, V.V. Ivanov$^{a}$, W.W. Jacobs$^{b}$,
     W. K\"{u}hn$^{h}$, A. Maggiora$^{d}$, M. Maggiora$^{d}$, 
     A. Manara$^{b,d}$, D. Panzieri$^{d}$, H. Pfaff$^{h}$,
     G. Piragino$^{d}$, G.B. Pontecorvo$^{a,d}$, A. Popov$^{a}$, 
     J. Ritman$^{h}$, P. Salabura$^{f}$, P. Senger$^{e}$, J. Stroth$^{i}$, 
     V. Tchalyshev$^{a}$, F. Tosello$^{d}$, S.E. Vigdor$^{b}$, G. Zosi$^{d}$   
     \\
     \textit{a}) JINR - Dubna;      \\
     \textit{b}) IUCF - Indiana;    \\
     \textit{c}) LNS - CEN Saclay;  \\
     \textit{d}) Dipartimento di Fisica Generale "A. Avogadro" \\
                 and INFN - Torino;                    \\
     \textit{e}) GSI - Darmstadt;                      \\
     \textit{f}) Institute of Physics - Krakow;        \\
     \textit{g}) Jagellonian University - Krakow;      \\
     \textit{h}) II Physikalisches Institut - Giessen; \\
     \textit{i}) Institut fur Kernphysik - Frankfurt;  \\
     \textit{l}) TRIUMF - Vancouver.}

\section{Introduction}

The study of strangeness producton in $\vec{p} p$ reactions is the 
central research project of the DISTO Collaboration~\cite{disto_ref} 
at Saturne.

Two major related physical questions:  
I)  the puzzling problem of the hyperon polarization;
II) the surprising large violation of the OZI rule in several reactions,
constitute the core of the DISTO program.

In this paper, I will briefly describe the present status of this two
questions and I will show some very preliminary results 
of the first production running.
 
\subsection{Hyperon polarization}

One of the most puzzling and persistent, since long time, spin effect was 
observed in inclusive hyperon production in collisions of unpolarized
hadron beams. A very significant polarization of the $\Lambda$-hyperon 
was discovered at Fermilab more than two decades ago~\cite{bunce},
in certain kinematical conditions.

This astonishing result was verified by many different experiments in the 
energy range from 10 up to 2000 GeV/c$^{2}$.
The $\Lambda$ polarization was observed along the normal direction to the 
$\Lambda$ production plane defined by the cross product of the incident 
beam direction with the produced hyperon direction 
\begin{displaymath}
\hat{n}_{\Lambda} = \frac
          {\mathbf{k}_{beam} \times \mathbf{k}_{\Lambda}  }
          {|\mathbf{k}_{beam} \times \mathbf{k}_{\Lambda}|}
\end{displaymath}
where $\mathbf{k}$ is the momentum of the projectile or the produced $\Lambda$
hyperon.

The $\Lambda$ weak decay into a proton and $\pi^{-}$ with a 64.1\% branching
can be used to extract the normal polarization from the 
parity-violating asymmetry of the decay proton having the emission angle 
$\theta^{*}$ with respect to $\hat{n}_{\Lambda}$ in the $\Lambda$ rest frame.
\begin{equation}
    \frac{dN}{d \cos \theta^{*}} = N_{0} (1 + \alpha P_{\Lambda} 
                           \cdot \cos\ \theta^{*})
  \label{pol_formula}
\end{equation}
where $\alpha$ is the measure of mixing of parities in the decay;
$\alpha = 0.642 \pm 0.013$.

The behavior of the polarization is usually expressed in terms of three   
kinematics variables: total C.M. energy $E^{*} = \sqrt{s} / 2$, 
Feynman's variable $x_{F} = p^{*}_{L} / p^{*}_{max}$, 
and transverse momentum $p_{T} = p \: \sin \! \phi$.

Based primarily on the huge $\Lambda$ polarization data presently available, 
which span the largest kinematic region, the polarization was considered 
consistent with the kinematic behavior summarized below: 

\begin{itemize}
  \item it is roughly independent of C.M. energy between about 10 to 2000 
        GeV/c$^{2}$; 
  \item it linearly increases with $p_{T}$ up to about 1 GeV/c. Above 1 GeV/c 
        the polarization is constant;
  \item it is compatible with 0 when $\overline{\Lambda}$'s are produced by a 
	proton beam but is not equal to 0 when they are produced by a 
	$\bar{\mathrm{p}}$ beam~\cite{banerj};
  \item it linearly increases with {\it x} up to $p_{T} = 1$ GeV/c; it is 
        independent of $p_{T}$ above;
  \item it is weakly dependent on the target type, and decreases with 
        increasing atomic weight~\cite{pondrom};
  \item it is positive when $\Sigma$ particles are produced but negative 
        for $\Lambda$ particles inclusive production.
\end{itemize}

A more complete review and the references to the original experiments
can be found in \Ref{heller}.

This behavior was generalized and extended to the polarization of
other hyperons and was thought to be a general behavior of polarization 
phenomenon.
However, recent data have cast great doubt on such a hasty conclusion. 

The fact that early experiments had shown $\overline{\Lambda}$ to be 
unpolarized, whereas, in the same kinematical region the $\Lambda$ 
was polarized, lent credence to the idea that polarization is a leading 
particle effect. This was supported by measurements~\cite{luk} showing 
the $\Omega^{-}$ to be unpolarized in this kinematical region. Two out of 
three quarks of the $\Lambda$ are the valence quarks of the incident proton
and the quark \textit{s} is picked-up from the sea. The $\Omega^{-}$
is composed of three strange valence quarks, it contains none of the 
valence quarks of the incident proton. 

But the recent measurements of the $\overline{\Xi}^{+}$ polarization by 
the E576 experiment at Fermilab~\cite{ho} show $\overline{\Xi}^{+}$ to be 
polarized by about the same amount as the $\Xi^{-}$.
Similar results are obtained by E761 experiment~\cite{morelos} which showed 
that $\overline{\Sigma}^{-}$ hyperons are produced in high energy collisions 
with a polarization of the same sign, though roughly half the magnitude of 
that of $\Sigma^{+}$. Moreover, for the first time, the $\Sigma^{+}$ 
polarization was observed to increase with $p_{T}$, achieve a maximum near 
$p_{T}$ = 1 GeV/c and then decrease.

This would indicate that the polarization of anti-hyperons is a common
phenomenon, and we should now turn our attention to why the
$\overline{\Lambda}$ and the $\Omega^{-}$ are not produced polarized.

It must be mentioned that the $\Lambda$ polarization data quoted until now
have been obtained in inclusive measurements, i.e. in reactions where only one
of the reaction products is measured. 

The directly produced  $\Lambda$'s cannot be distinguished from those 
coming from the decay of other hyperons like $\Sigma^{0}$  
\mbox{( $\Sigma^{0} \rightarrow \Lambda \gamma$ )} or S = -1 resonances 
(Y$^{*}$) or nucleonic resonances (N$^{*}$) which decayed strongly to the 
measured final $\Lambda$ particle. 

The important role played by the hyperon resonance $\Sigma^{*}$(1385) and  
the mesonic resonance $K^{*}$(892) in the $\Lambda$ polarization was 
pointed out by recent results~\cite{drutskoi}. 
They reveal how the new measurements, which are able to disentangle the 
contribution to the polarization coming from all the possibles sources, are 
required to clarify the present complexity and richness of the experimental 
scenario.

To explain these puzzling data, several theoretical attempts have been made 
during the last decades~\cite{ander,degran,szwed,dharma,heller78,boros0,
troshin,prepa,hama,soffer,laget}. 

The proposed models span from the ``QCD inspired models'' to the 
``boson-exchange models''
with a wide range of different flavors. However, they have had
little real exhaustive predictive power and none of them deal, for instance, 
with the polarization of antihyperons (with the exception of the 
hydro-dynamical model~\cite{hama}).

Two new ``dynamical QCD models''~\cite{boros0,troshin} seem to be more
promising when used to predict other polarization observables (see below).

A more exhaustive review can be found in the Kroll~\cite{kroll} or  
Soffer~\cite{soffer1} articles. Although rather old (they do not account for 
the most recent models) and written 
before the polarization of the $\overline{\Xi}^{+}$ and 
$\overline{\Sigma}^{-}$ was discovered, they are, in my knowledge, 
the best theoretical compilations yet.

\subsection{Spin observables in hyperon production}

The availability of polarized proton beams allows the study
of a new set of spin observables other than polarization $P$, viz.,
analyzing power $A_{N}$ and depolarization $D_{NN}$.
\begin{eqnarray*}
 A_{N} & \equiv & \frac{\sigma(\uparrow) - \sigma(\downarrow)}
                   {\sigma(\uparrow) + \sigma(\downarrow)} 
                                                       \label{an_formula}\\
 D_{NN} & \equiv & \frac
                      {\sigma(\uparrow\uparrow)+\sigma(\downarrow\downarrow) 
                      -\sigma(\uparrow\downarrow)-\sigma(\downarrow\uparrow)}
                      {\sigma(\uparrow\uparrow)+\sigma(\downarrow\downarrow)
                      +\sigma(\uparrow\downarrow)+\sigma(\downarrow\uparrow)}
                                                        \label{dnn_formula}
\end{eqnarray*}
where $\sigma$ is the pure spin cross section, the first arrow refers to 
the beam polarization direction and the second one refers to the measured 
hyperon polarization~\cite{aip_conf_42}.

They are a generalization of the Wolfenstein spin-rotation parameters
for proton-proton elastic scattering~\cite{wolfenstein} and are
theoretically defined as a ratio of cross sections. So all of the
normalizations used to calculate cross sections cancel, leaving only the
key parameter of the interaction.

Measurements of these parameters, particularly $D_{NN}$ for which the 
prediction is parameter-free in some models, provide crucial tests of the
models and its assumptions that the process can be treated at the quark
level.

Due to the lack of good quality polarized beams, the experimental scenario
is rather poor compared to the inclusive hyperon polarization data.
Only three measurements are available up to now: at 6 GeV/c~\cite{lesnik},
at 13.3 and 18.5 GeV/c~\cite{bonner} and 200 GeV/c~\cite{bravar}.

The common feature of these inclusive experimental data, when compared in the 
same kinematical range, is a relevant asymmetry and a substantial spin 
transfer as large as 30\% in the beam fragmentation region. No data
are available in the target fragmentation region due to the limited
acceptance of these experiments. 
 
A recently proposed model~\cite{boros}, based on the idea of rotating 
constituents in polarized proton, in the direction suggested by the results 
of deep inelastic experiments~\cite{ashman}, is fairly successful 
in accounting the observed $A_{N}$ behavior.
The $D_{NN}$ trend is also qualitatively reproduced by this model.

\subsection{Intrinsic strangeness content of the nucleon}

The idea that strange quarks may reside in the nucleon was pointed out by
J. Ellis, E. Gabathuler and M. Karliner~\cite{ellis0} to provide an
explanation for several experimental puzzles. 

The first one was the controversial problem of the SU(3) chiral symmetry 
breaking operator measured through the $\pi$-N $\sigma$-term~\cite{cheng}. 
This term is a factor 2 higher than the one expected from the Gell-Mann-Okubo
mass formula and the assumption $<\mathrm{p}|\bar{s}s|\mathrm{p}> = 0$.
This discrepancy is explained in recent lattice QCD calculations~\cite{dong}
assuming a sizeable contents of sea $\bar{s}s$ pairs inside the nucleon.

The second point is the famous result of deep inelastic scattering 
measurements~\cite{ashman}, which indicate that 
$\Delta s = -0.10 \pm 0.03$, where $\Delta s$ is the fraction of spin
carried by strange quarks and antiquarks. The minus sign means that the 
strange $q$ and $\bar{q}$ have a net polarization opposite to the direction 
of the nucleon spin.  

Moreover, the high $\overline{K} K$ yield in $\bar{p} p$ annihilation at rest, 
the backward peak in $\bar{p} p \rightarrow K^{-} K^{+}$
reaction at $p = 0.5 GeV/c$, and the anomalous high cross section
$\sigma(\bar{p} p \rightarrow \phi \phi)$ can be easily accommodated
in the intrinsic strangeness model of the nucleon~\cite{ellis}.

The last puzzle is the violation of the Okubo-Zweig-Iizuka 
(O.Z.I.) rule~\cite{okubo} experimentally observed in several reactions.

In the naive quark model, the O.Z.I. rule put limits on 
the possible schemes of quark rearrangement processes that lead to meson 
production. 

According to this rule, it was predicted~\cite{lipkin} that the possible
values of 
\begin{equation}
	R = \frac{\sigma(A + B \rightarrow \phi X)}
		 {\sigma(A + B \rightarrow \omega X)}
  \label{phi_omega_ratio}
\end{equation}
are $R = 4.2 \times 10^{-3}$ using the quadratic Gell-Mann-Okubo mass formula
or  $R = 0.15 \times 10^{-3}$ using the linear mass formula.

\begin{figure}
      \centering \includegraphics[width=0.48\textwidth]
                                 {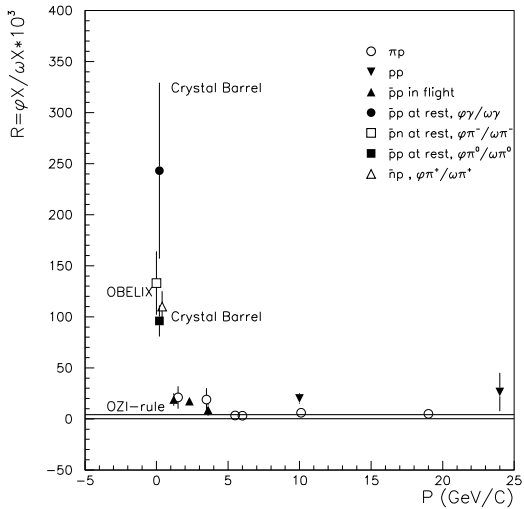}
      \caption{    Ratios $R = \phi X / \omega X$ in different reactions at
                   increasing momenta p.
                   The horizontal line is the theoretical prediction 
                   from quadratic Gell-Mann-Okubo mass formula
                   $R = 4.2 \cdot 10^{-3}$. See Ref.~\protect\cite{ellis}
                   for data and extensive figure caption.}
      \label{fig:ozi_vi}
\end{figure}

The experimental values of $R$, however, stay typically in
the range $(10 \div 20) \times 10^{-3}$. From this disagreement between 
experiment and theory a semi-empirical rule was given that 
implies~\cite{ellis} that the O.Z.I. rule is generally violated at least at 
the level of 10\%. As shown in \Fig{fig:ozi_vi} a much larger apparent 
violation was found in $\bar{p} p$ annihilation at rest~\cite{obelix}.

It was suggested~\cite{ellis0}~\cite{ellis2} that this strong violation of 
the O.Z.I. rule could be explained with the existence
of an admixture of a $\bar{s}s$ quark pair in the nucleon even at large
distances. In \Ref{ellis} it is shown that the 
amount of the admixture needed to accommodate the data is quite small. 
Moreover, it was shown~\cite{ellis3}, that the strange quark pairs are 
polarized in the opposite direction to the nucleon spin.

The study of the $\phi$-meson production is a particularly 
sensitive test of the validity of the O.Z.I. rule because the $\phi$ is 
almost a pure $\bar{s}s$ state, containing just a small admixture of 
$\bar{u}u + \bar{d}d$.

An interesting consequence of this is the possible link between the
polarization of the nucleon strange sea and the different yields of the
$\phi$-meson production from different spin states of the 
$\overline{\mathrm{N}}$N system and from different annihilation 
channels. Remembering that the intrinsic spin of the $\phi$ 
meson is $J = 1$, one can expect a maximum enhancement of $\phi$ production
in the $^{3}S_{1}$ channel where the spins of strange quarks and anti-strange 
quarks are parallel.

Among the possible checks~\cite{ellis} of this model, the most straightforward 
one would be to measure the $\phi$ production rate in the reaction 
$\vec{p} \vec{p} \rightarrow p p \phi$.
If the aforesaid assumptions are correct~\cite{rekalo}, the $\phi$ rate should 
be maximal when the spin of the beam and the spin of the target are oriented 
parallel.

Using the polarized beam of Saturne, DISTO could make a first step in that 
direction, as we will allow to absolute $\phi$ production rate 
and its dependence on the beam polarization. 
A comparison with the $\omega$ production rates in the same kinematical range,
measured in the same experimental apparatus, can also be made. 
This will be the first attempt to measure the $\phi$ production 
at threshold because there are only two measurements at 10 and 24 GeV/c 
(see \Fig{fig:ozi_vi}) far from the threshold.

\section{The DISTO program}

The DISTO experiment~\cite{disto_ref} was specifically designed to study the 
associated $\Lambda$ and $\Sigma^{0}$ production
\begin{equation}
  \vec{p} p \rightarrow p K^{+} Y \tag{Y = $\Lambda$, $\Sigma^{0}$, $Y^{*}$}
\end{equation}
and the vector meson production 
\begin{equation}
	\vec{p} p \rightarrow p p V_{m} \tag{$V_{m} = \phi, \omega$}
\end{equation}
at 2.85 GeV, the maximum usable energy of Saturne, and 2.5 GeV for hyperon 
production.

The experimental set-up (see \Fig{fig:exapp}) is designed to track the 
four charged products signaling hyperon or vector meson production 
through a strong magnetic field.
The measurements of the angles and momenta of the $\Lambda$ (or $\phi$) decay  
products, of the primary proton and of the associated kaon (or proton) will
allow a complete kinematical reconstruction of the missing mass of the 
reaction products. Only in this way the contributions to the $\Lambda$ 
polarization coming from different sources can be disentangled.
For the $\Sigma^{0}$ production, only the photon from the (100\% branch)
$\Sigma^{0} \rightarrow \Lambda \gamma$ decay is missing.
For the $\omega$ production, only the $\pi^{0}$ from the (88.8\% branch)
$\pi^{0} \rightarrow \pi^{+} \pi^{-} \pi^{0}$ decay is missing.

Taking advantage of the high quality of the polarized beam produced by Saturne,
the DISTO collaboration plans to:

\begin{itemize}
\item measure the differential cross-sections\\
         $d\sigma / d\Omega$ for $\Lambda$, $\Sigma^{0}$, and $Y^{*}$ 
         productions;
\item measure the polarization $P$ of the hyperons produced;
\item study the dependence on the beam polarization of these observables
	getting the analyzing power $A_{N}$, and the depolarization
	parameter $D_{NN}$;
\item study the relationship between these observables and the N$^{*}$ 
	and Y$^{*}$ ($S = -1$) resonances; 
\item measure the differential cross-section and the analyzing power $A_{N}$
	of the reaction $p p \rightarrow p p \phi$ near the production 
	threshold and the $\phi$/$\omega$ ratio (\Eq{phi_omega_ratio}).
\end{itemize}

This experiment is the first attempt to carry out a complete study 
(including the spin) of the reaction mechanisms through an exclusive 
measurement.
The measurement of $D_{NN}$, simultaneously for $\Lambda$ and $\Sigma^{0}$
production, is expected to provide an especially strong constraint on 
various theoretical models.

\section{The DISTO set-up}

\begin{figure}
     \centering \includegraphics[width=0.48\textwidth]
                                 {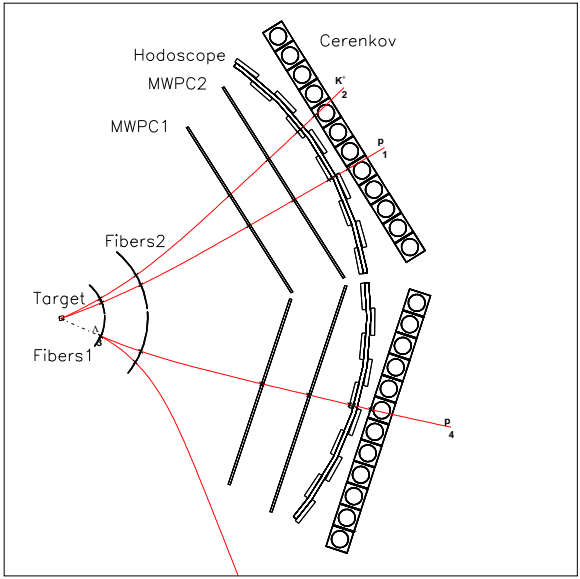}
        \caption{    Layout of the DISTO experimental set-up, shown in 
                     plan view. A simulated 
                     $\vec{p} p \rightarrow p K^{+} \vec{Y}$ event is also 
                     shown. The detector assembly is described in the text.}
        \label{fig:exapp}
\end{figure}

The experimental setup, shown in \Fig{fig:exapp}, includes the magnet 
S170 from CERN which provides a 
maximal magnetic field of 14.7~kGauss, an angular acceptance 
$\Delta \theta = \pm 120^{\circ}$ in the horizontal plane and 
$\Delta \phi = \pm 20^{\circ}$ in the vertical plane.

The detector array is designed to cover a dip angle of $\pm15.5^{\circ}$
and a scattering angle of $45^{\circ}$ on both sides of the curved beam 
trajectory.

The tracking detectors comprise two left-right pairs of semi-cylindrical
scintillating fiber chambers (two stereo layers, \emph{u-v} planes, and one 
horizontal \emph{y} plane, of 1 mm square fibers); similarly two pairs of 
\emph{x-u-v} planar multi-wire proportional chambers will be mounted at the 
edge of the magnet poles.

Located radially at about 140~cm from the unpolarized 2-cm thick liquid 
hydrogen target, a scintillator hodoscope records particle multiplicities,
allows p vs K$^{+}$ time-of-flight particle identification, and provides a
sample of $\vec{\mathrm{p}}$p elastic scattering events to monitor the beam
polarization and intensity.

Finally, behind the hodoscope, a vertically segmented Cerenkov counter gives 
information on the velocity of crossing particles 
making easier the separation between pions, kaons and protons on the whole
spectrum of particles produced. 

The trigger selects events with at least four charged prongs
within the detector acceptance, using the hit multiplicity information
given by the scintillation fiber detectors and by the hodoscope.

Although the trigger rate sustained by data acquisition system is 
$\geq$ 10~KHz per spill (0.5 s), it is kept around 3~KHz events per spill by
beam intensity and trigger logic, in order to have a dead-time of 
roughly 10\%.

\section{Preliminary results}

The DISTO detector assembly has been completed during July 1996.
The first production running with a fully operational detectors system,
level 1 trigger and data acquisition was done during past November 1996.
During 10 days of beam around $500 \cdot 10^{6}$ triggers have been recorded.

\begin{figure}
      \centering \includegraphics[width=0.48\textwidth]
                                 {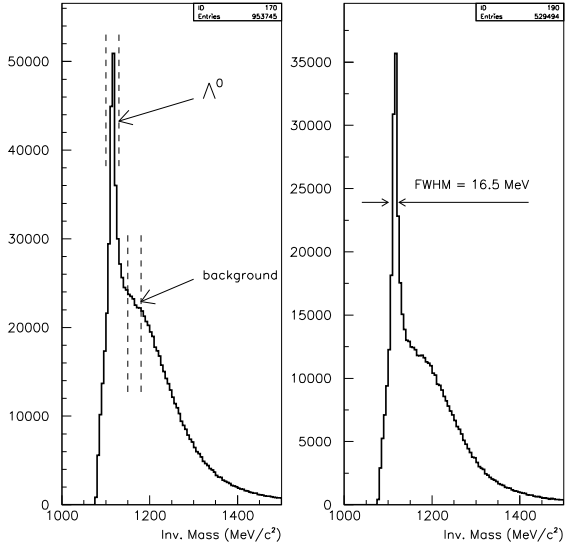}
         \caption{    $p \pi^{-}$ invariant mass spectra after the standard
                      software cuts. The present conservative approach leaves
                      significant background in these spectra. 
                      Left: no cuts on 4-track Missing Mass. The $\Lambda$
                      and background cuts used in the subsequent analysis
                      are also shown. 
                      Right: Improved signal / background ratio with the 
                      additional cut:
                      $-0.10 \geq MM(4-track) \geq 0.08 GeV^{2}$}
         \label{fig:inv_mass}
\end{figure}

\begin{figure}
      \centering \includegraphics[width=0.48\textwidth]
                                 {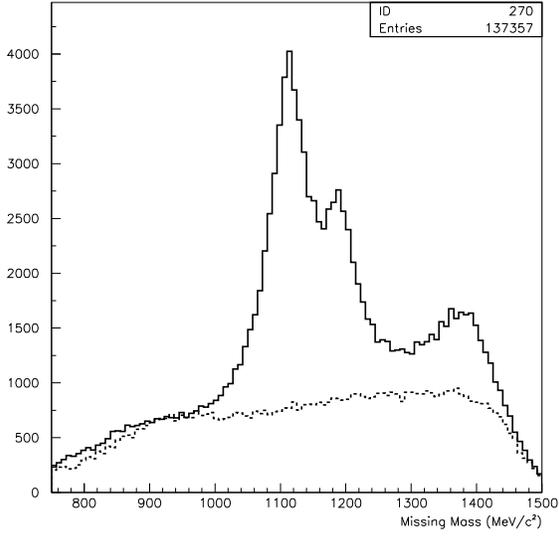}
         \caption{    $p K^{+}$ missing mass spectra for $\Lambda$ 
                      (solid line) and background events (dashed
                      line) (see Fig.~\protect\ref{fig:inv_mass}). 
                      The background is clearly not resonant.}
         \label{fig:miss_mass1}
\end{figure}

\begin{figure}
      \centering \includegraphics[width=0.48\textwidth]
                                 {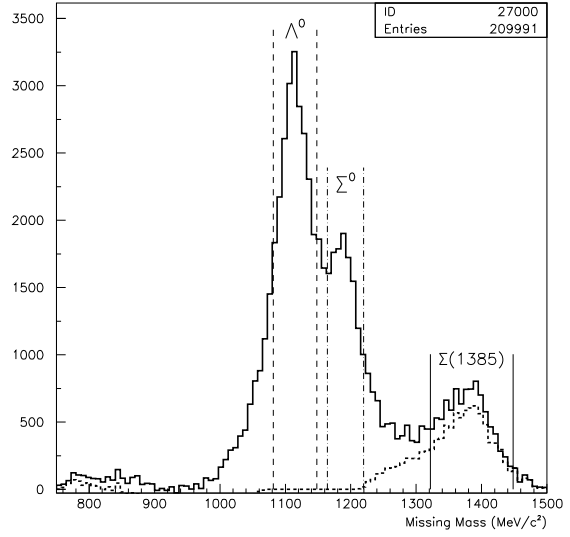}
         \caption{    Background subtracted $p K^{+}$ missing mass spectra. 
                      The prominent peaks are $\Lambda$(1115), 
                      $\Sigma^{0}$(1192), and 
                      $\Sigma$(1385). 
                      Dashed line: missing $MM(4-track) \geq \pi^{0}$ cut; 
                      the cut eliminates
                      the $\Lambda$ and $\Sigma$(1192) peaks.}
         \label{fig:miss_mass2}
\end{figure}

\begin{figure}
      \centering \includegraphics[width=0.48\textwidth]
                                 {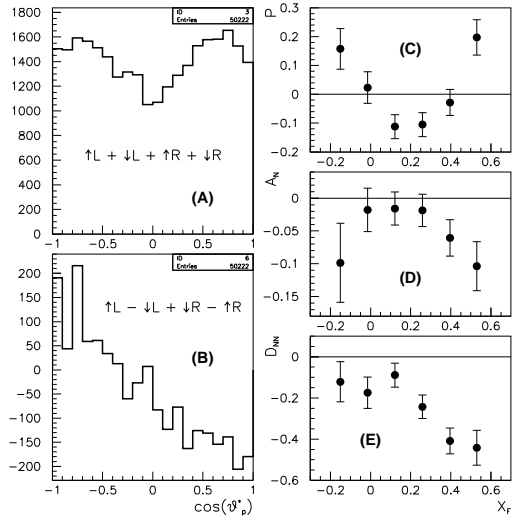}
         \caption{    Background subtracted decay distribution for 
                      $\Lambda$(1115).  L and R refer to $\Lambda$ produced
                      to the beam left or right; the arrows refer to the                             beam polarization direction. 
                      A) $P_{\Lambda}$: integrated 
                         ($\vec{k}_{p_{\Lambda}} \cdot \hat{n}$) distribution 
                         in the $\Lambda$ rest frame. 
                      B) $D_{NN}$: integrated spin difference distribution. 
                      C), D), E): polarization, analyzing power and 
                      depolarization parameter vs. $X_{F}$}
         \label{fig:costeta}
\end{figure}

\begin{figure}
      \centering \includegraphics[width=0.48\textwidth]{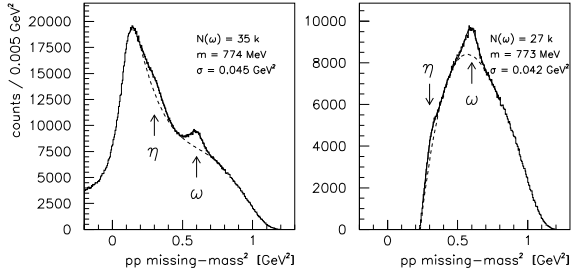}
         \caption{    pp missing mass spectra. 
                      Left: conservative cut; 
                      Right: missed $\pi^{0}$ cut}
         \label{fig:omega_mimass}
\end{figure}

\begin{figure}
      \centering \includegraphics[width=0.48\textwidth]{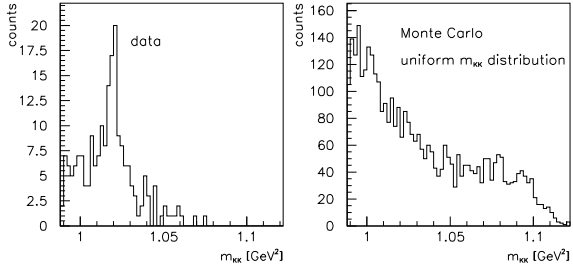}
         \caption{    KK missing mass spectra. 
                      Left: The $\phi$ peak is clearly shown. cut: 4-track
                      invariant mass = 0
                      Right: phase-space simulation with the same cuts.}
         \label{fig:pi_kk_mass}
\end{figure}

\begin{figure}
      \centering \includegraphics[width=0.48\textwidth]{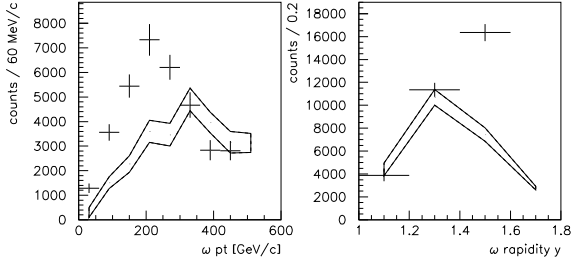}
         \caption{    kinematical correlation of events in the $\omega$
                      missing-mass region.
                      solid line bands: expected distribution from a 
                      phase-space Monte Carlo simulations}
         \label{fig:th_omega_y_pt}
\end{figure}

\begin{figure}
      \centering \includegraphics[width=0.48\textwidth]{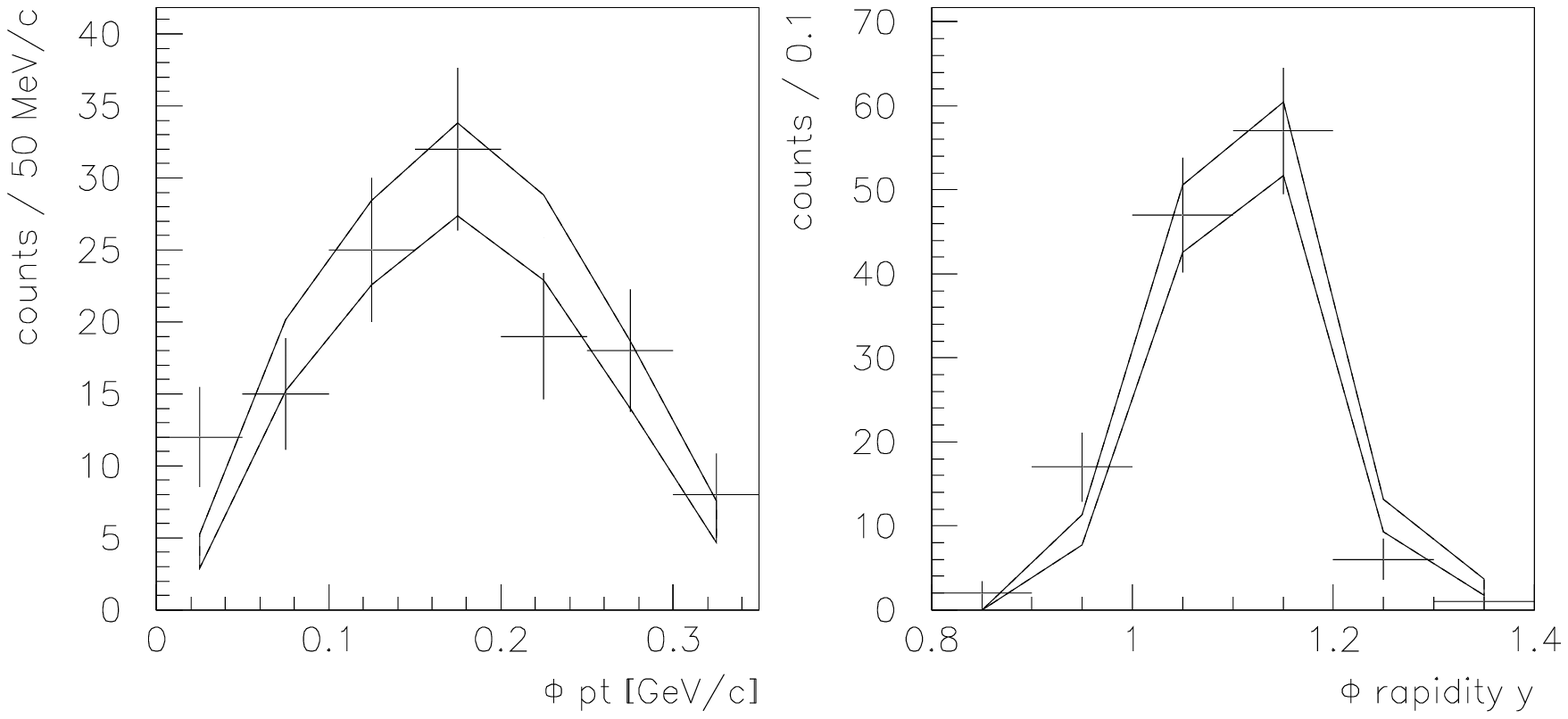}
         \caption{    Kinematical distribution of $K^{+} K^{-}$ pair in the 
                      mass region m$_\phi \pm$ 15 MeV.
                      solid line bands: expected distribution from a 
                      phase-space Monte Carlo simulations}
         \label{fig:th_phi_y_pt}
\end{figure}

Some preliminary results of the November '96 and May '97 production
runs (around 600 runs) are shown in 
Fig.~\ref{fig:inv_mass}-\ref{fig:th_phi_y_pt}. 

\Fig{fig:inv_mass} show the $p \pi^{-}$ invariant mass spectra left
by the standard cuts. Each of the cuts has been made sufficiently conservative
to remove essentially no events in the $\Lambda$ peak.

This conservative approach leaves significant background in these spectra.
In what follows, we often subtract background from subsequent spectra by
means of a cut on invariant mass outside the peak.
This background is basically flat and not resonant (see 
\Fig{fig:miss_mass1}) so this technique is essentially bias free. 

Making cuts a bit tighter or adding additional cuts 
(\Fig{fig:inv_mass} right) would reduce the background by a large factor,
at the expense of 10-20\% of the pKY events.

\Fig{fig:miss_mass1} shows the missing mass spectra reconstructed from
the reaction products identified as the primary proton and kaon, for events
within the $\Lambda$ invariant mass peak and within the background gate.

In the background subtracted spectrum shown in \Fig{fig:miss_mass2},
we see clear peaks for the $\Lambda$(1115), the $\Sigma$(1192), and the
$\Sigma$(1385). 
Also shown is the background subtracted spectrum gated by a cut requiring
the overall missing mass to exceed 0.01 GeV$^{2}$, in order to select
only events missing a pion, a pion and gamma, two pions etc.
This cut essentially eliminates the  $\Lambda$ and  $\Sigma$(1192) peaks,
but leaves $\Sigma$(1385), which decays to $\Lambda \pi$ or $\Sigma \pi$.

For the analysis of polarization results, we separate $\Lambda$,  
$\Sigma$(1192) and $\Sigma$(1385) events by the purposely narrow cuts
shown in \Fig{fig:miss_mass2}.

Some preliminary polarization results are shown in \Fig{fig:costeta}.
The histogram (A) shows the decay angle distribution summed for $\Lambda$'s
left and right and for spin up and down, and summed over $P_{T}$ and
$X_{F}$. The hyperon polarization,
given by \Eq{pol_formula}, would show up via a systematic fore-aft
asymmetry in this spectrum, which is not seen in this distribtion
not corrected for geometrical acceptance.

In contrast histogram (B) shows the spin-difference spectrum relevant
to the polarization transfer in 
$\vec{p} p \rightarrow p K^{+} \vec{\Lambda}$.
Here a sizeable fore-aft asymmetry is seen indicating a large value
of $D_{NN}$, even when the yields are summed over the full acceptance,
$P_{T}$, and $X_{F}$. $D_{NN}$ being a ratio of cross-sections, the
geometrical acceptance corrections are not relevant, at least in the first
order. 

Histograms (C), (D), and (E) show the $\Lambda$ polarization (P), analyzing
power (A$_{N}$) and depolarization parameter (D$_{NN}$) as a function of 
the Feynman scaling variable $X_{F}$, summed over $P_{T}$.

These preliminary results indicate a substantial polarization, analyzing
power and depolarization parameter, as large as 40\%, in the beam 
fragmentation region, i.e for $X_{F}$ large and positive.

For the first time it is also possible to study the polarization observables
approaching the target fragmentation region. 
Although the error bars are relevant with the present statistics, clear 
indication of polarization and analyzing power 
behavior similar to that observed in the beam fragmentation region can be 
shown. Viceversa $D_{NN}$ is compatible with 0 in this region. 

Fig.~\ref{fig:omega_mimass}-\ref{fig:th_phi_y_pt} show the first preliminary
results on vector meson production.

\Fig{fig:omega_mimass} shows the $p p$ missing mass with two different
cuts. 
Left:  $m_{\pi \pi}^{2} \leq 0.41$ GeV$^{2}$;
right: $m_{cm - pp}^{2} - m_{\pi \pi}^{2} \leq 0.15$ GeV$^{2}$.

By fitting a Gaussian function to the signal, the number of events of the type
$ p p \omega$ can be estimated. The background is fitted with a third order
polynomial function. The signal from $p p \eta$ is also taken into account by
adding an additional Gaussian function in the fit. 

\Fig{fig:pi_kk_mass} shows the $K^{+} K^{-}$ pair invariant mass, 
compared to a Monte Carlo simulation with uniform $m_{KK}$ distribution
and a phase space distribution for the system $p, p, K, K$.

The two kaons are identified using the water Cerenkov pulse height and
several geometrical and kinematical boundary conditions, mainly
$m_{cm - pp}^{2} - m_{K K}^{2} \approx 0$.

\Fig{fig:th_omega_y_pt} and \ref{fig:th_phi_y_pt} show the kinematical
distribution of the $\omega$ and $\phi$ mesons as a function of $P_{T}$
and rapidity $Y$.

For comparison the solid line bands show the distribution from a Monte Carlo
simulation using 3-body phase space.

Although there is no significant deviation from phase space in $\phi$
production, the $\omega$ shows a relevant difference. This means a careful
simulation is required for doing accurate acceptance corrections.

\section{Conclusions}

The good preliminary results shown in the previous section, obtained with
only 1/5 of the final statistics foreseen at 2.85 GeV, allow us to say
that DISTO experiment could make an essential contribution to clarifying
the present puzzle of polarization phenomena and OZI rule and, in turn, 
the role of the strangeness in the nucleon.

However some ``caveat'' have to be expressed for a correct interpretation
of the results presented:
I) the beam polarization was determined on-line by our polarimeter.
        A more precise determination of the beam polarization is in progress
        by off-line analysis. 
II) no acceptance corrections are included as well as no dead-time
        corrections.

In spite of these reserves very important results are already shown by our
data:
\begin{enumerate}
  \item $\phi$ and $\omega$ production is clearly shown in our data. Once more
        statistics will be available and the acceptance corrections under 
        control, the determination of the relative production rates will be
        possible and an important contribution to the OZI rule puzzle
        will be given. The determination of the analyzing power for 
        two vector mesons will also be possible. 
  \item an important result, still observable even if integrated on the whole
        acceptance of DISTO set-up, is the large value of the depolarization
        parameter $D_{NN}$ for direct $\Lambda$ production. This indicates
        a transfer of polarization, during the transition, from the beam to
        the $\Lambda$. At the same time P is small when summed over $P_{T}$
        and $X_{F}$ but increases with $X_{F}$ in the beam fragmentation 
        region. This result is in agreement with those obtained in
        inclusive production at 6 GeV/c~\cite{lesnik} and 
        200 GeV/c~\cite{bravar}. \\
        Once the full statistic will be available we can study the dependence
        of the spin observables on the different kinematical parameters
        for the $\Lambda$, $\Sigma$(1192), and $\Sigma$(1385).
  \item the measurement at 2.5 GeV, planned during this year, will allow
        to understand the reaction mechanism, by comparing the polarization 
        observables in an energy regime where
        the pion exchange is more favorite than the kaon exchange.
\end{enumerate}

At the end of this year the Saturne accelerator will be definitively closed,
depriving the physicist community of a high quality and unique tool to
study the fundamental interactions.


\end{document}